\documentclass[amsmath,amssymb,epsfig,aps,pra,twocolumn]{revtex4-1}
\usepackage{amsmath}
\usepackage{epsfig}
\usepackage{graphicx}
\usepackage{color}
\usepackage{amssymb}
\usepackage{epstopdf}
\begin{document} 
\title{Scattering of cold  $^4$He on $^4$He$-^{6,7}$Li and $^4$He$-^{23}$Na molecules} 
\author{M. A. Shalchi$^1$, A. Delfino$^{2}$, T. Frederico$^{3}$ and Lauro Tomio$^{1,3}$\\ }
\affiliation{
$^{1}$ Instituto de F\'\i sica Te\'orica, Universidade Estadual Paulista, 01405-900 S\~{a}o Paulo, Brasil.\\
$^{2}$ Instituto de F\'\i sica, Universidade Federal Fluminense, 24210-310 Niter\'oi, RJ, Brasil.\\
$^{3}$ Instituto Tecnol\'ogico de Aeron\'autica, DCTA, 12228-900 S\~ao 
Jos\'e dos Campos, Brasil.
}
\date{\today}
\begin{abstract}
We predict $s-$wave elastic cross-sections $\sigma$ for low-energy atom-molecule collisions with kinetic energies 
up to 40 mK, for  the $^4$He collision with weakly bound diatomic molecules  formed by $^4$He with $^7$Li, $^6$Li and $^{23}$Na. 
Our scattering calculations are performed by using diatomic and triatomic molecular binding energies obtained 
from several available realistic models as input in a renormalized zero-range model, as well as a finite-range 
one-term separable potential in order to quantify the relevance of range corrections to our predictions. 
Of particular relevance for possible experimental realization, we show the occurrence of a zero in 
$\sigma$ for the collision  of cold $^4$He on $^4$He$-^{23}$Na molecule below 20 mK.
Also our results for the elastic collision $^4$He on $^4$He$-^{6,7}$Li molecules suggest that $\sigma$  varies considerably for the 
realistic models studied.
As the  chosen molecules are weakly bound and the scattering energies are very low, our results are interpreted on 
the light of the Efimov physics, which explains the model independent and robustness of our predictions, despite some 
sensitivity on the potential range.
\end{abstract}

\maketitle

\section{Introduction} 
The Efimov effect~\cite{Efimov} is a peculiar pure quantum-mechanical effect, expected to occur in a three-body quantum systems,
manifested by an increasing number of three-body bound states when the absolute value of the scattering length of a 
two-body subsystem is approaching to infinite. 
This effect have a long tradition of studies in nuclear physics context, being sometime mentioned as {\it Thomas-Efimov} 
effect~\cite{Thomas-Efimov}, considering its relation to another property of the three-body Schr\"odinger formalism noticed by 
Thomas~\cite{thomas} in 1935, when investigating the origin of the nuclear forces between nucleons.
By considering a non-relativistic two-boson interaction supporting one bound state, he observed 
that the three-body ground-state will collapse to  $-\infty$ in the limit when the range of the interaction is zero. 
This observation was essential for the first conclusions on the range of about one femtometer of the nuclear 
forces. Besides the fact that the initial investigations on Efimov states in nuclear physics have been limited
to theoretical approaches not experimentally realizable (as the two-body interaction is fixed),
we should notice some theoretical efforts in given evidence that some well-known states could eventually
be considered as manifestations of Efimov states, by considering the behavior of such states when varying 
the potential parameters such that the two-body interaction is driven to the unitary limit. 
In particular, this is the case of the original proposal that the virtual state of the $s-$wave spin doublet 
trinucleon system is an Efimov state~\cite{1982-PRC}.
The interest on verifying manifestations of Efimov physics in nuclear physics came much later with the discovery
of exotic nuclei~\cite{tanihata}  having two neutrons far apart from a core~\cite{Jensen,Bhasin,1997-Amorim}.
Since then, extensive investigations on universal aspects of light halo nuclei are available, in the context of 
Efimov physics, which can be traced by several reviews. For that, we can mention the 
Refs.~\cite{2002-Bedaque,2006-Braaten,2009-Epelbaum,2012-review,2017-Hammer}, 
in which the updated review on halo-nuclei description of Ref.~\cite{2017-Hammer} is exploring the effective field theory approach. 
In view of the limitations to observe indications of Efimov effect coming from nuclear physics aspects, most of the 
initial theoretical studies on Efimov states have been considered three-atom systems, by using realistic interatomic 
interactions~\cite{1977lim,1978lim,1983lim,1998Yuan,1984CG,1999-Amorim}. 
The trimer of $^{4}$He, due to the very weak binding of the corresponding dimer, was
long-time predicted in 1977 to present an Efimov state in Ref.~\cite{1977lim}, on the basis of a three-body 
calculation in momentum space using Faddeev formalism. Their investigation was followed by several other 
related works done in the same period~\cite{1978lim,1983lim}.  
Later on, in another independent work within the Faddeev scheme, Cornelius and Gl\"ockle~\cite{1984CG} confirmed
the existence of two bound states for the $^{4}$He trimer, with the weakly-bound excited state having the property of 
an Efimov state.
The existence of the weakly-bound excited state in helium, stablished in Ref.~\cite{1984CG}, also proved to be a 
good test for the predicting power of the scaling approach presented in Ref.~\cite{1999-Amorim}
(essentially the same result is obtained). The search of Efimov states in such a 
system~\cite{1985Huber,1996Esry,1997Kolganova,2000Delfino,2000Gianturco,2002Gianturco,2012Roudnev,2014Wu,2017Kolganova} 
has been motivated by the remarkable small 
binding energy of the $\,^{4}$He dimer: $B_{^4\rm{He}_2} =$ 1.31 mK~\cite{2000-Hedimer}. 
Finally, in 2015, it was reported in Ref.~\cite{2015-kunitski}, the 
experimental observation of this long-time predicted Efimov state. 
The experimental success in verifying such long-time theoretical prediction, together with the results of previous
experimental investigations of Efimov physics in coldatom laboratories~\cite{bec-exp-rev}, which are extended
to mixed atomic-molecular combinations~\cite{bec-exp-mix},  became highly motivating 
for more deeper theoretical studies with single or mixed atomic 
species~\cite{BringasPRA04,2010wang,2011levinsen,2012Garrido,2014blume,2018Shalchi}.
Quite remarkable are the advances in the laboratory techniques, such that one can even consider the possibility to
alter the two-body interaction by using Feshbach resonance mechanisms (originally proposed in the nuclear physics 
context)~\cite{avaria}. In ultracold atom experiments, the possibility of changing the two-body scattering length was 
shown that can alter in an essential way the balance between the non-linear first few terms of the mean-field description 
which is modeling the atomic Bose-Einstein condensation~\cite{gammal}.  

In the present work, by following previous studies on triatomic molecules involving the Helium atom, in particular 
considering available results reported in Ref.~\cite{2017Suno} for realistic interactions, we are studying the 
cold atom-dimer elastic collision. Our study is focused in the cases where the three-body system is composed by 
a mixture with $^{4}$He and another atomic species chosen as being $^{6}$Li, $^{7}$Li, and $^{23}$Na. In all the 
cases, we assume $^{4}$He as the colliding particle with the dimer formed by the remaining two-body subsystem.
For the present study, we consider the Faddeev formalism using finite-range separable two-body interactions, 
as well as the renormalized zero-range (ZR) model~\cite{1997-Amorim}. The main observables that we are concerned
as relevant for possible experimental investigations, are the $s-$wave phase-shifts and the elastic $s-$wave 
cross-section for different colliding and dimer energies. 
In order to help us the analysis of the $s-$wave elastic scattering amplitude, the results for the 
absorption parameter are also presented in some relevant cases.

As we are concerned with relatively low kinetic colliding energies, with the lowest partial wave being more relevant 
for the Efimov physics, we focus our study on the $s-$wave contribution to the total cross-section. 
The corresponding contributions due to higher partial waves, such as from  $p-$ and $d-$waves, which should 
appear for increasing kinetic energies, are left to be explored in a future related investigation. 
However, as it will be shown here, the more interesting outcome is verified for kinetic energies 
where $s-$wave is expected to dominate.

In the next section, we present the formalism. The main results with corresponding discussion are given in 
section III. In section IV we have our final remarks and conclusions.

\section{Faddeev Three-Body Formalism}
In the present section, we fix our notation and include the standard formalism for the elastic scattering amplitude of a
particle $\alpha$ colliding to a dimer $(\alpha\beta)$, which is formed by the same particle $\alpha$ with another 
particle $\beta$. For convenience, as explained in our introduction, we choose $\alpha$ as the $^4$He atom, with 
$\beta$ being $^7$Li,  $^6$Li, or $^{23}$Na. In the following formalism, we are always considering that the three-body 
system $(\alpha\alpha\beta)$, as well as the subsystems $(\alpha\beta)$ and $(\alpha\alpha)$ are bound, such that we can 
take advantage of the corresponding available data as inputs coming from different realistic models, as well as from experimental 
considerations.
Therefore, everywhere along this presentation we are assuming as fixed the  $^4$He$_2$ binding energy 
and corresponding scattering length, such that $E_{\alpha\alpha}=-B_{\alpha\alpha}=-1.31$mK
and $a_{\alpha\alpha}=$100\AA.
The other input binding energies are obtained from specific 
models, which will be discussed. In particular, we should noticed the good agreement among most the realistic models 
on the other dimer binding energies ${\alpha\beta}$, such that the discrepancies coming from model results are mainly 
verified for the respective three-body energies. 
 
In the formalism, following Ref.~\cite{2017Shalchi}, 
we assume units such that  $\hbar=1$ (with energies given in mK), with $m\equiv m_\alpha = m_{^4{\rm He}}$ and 
a mass ratio which is defined by $A\equiv m_\beta/m_\alpha$, such that
 $\mu_{\alpha\alpha}=m/2$ and $\mu_{\alpha\beta}=Am/(A+1)$ are the reduced masses for the $\alpha\alpha$ and 
 $\alpha\beta$ subsystems, respectively, with the corresponding three-body reduced masses given
by $\mu_{\alpha(\alpha\beta)}=m(A+1)/(A+2)$ for the $\alpha- (\alpha\beta)$; and 
$\mu_{\beta(\alpha\alpha)}=m(2A)/(A+2)$ for the $\beta- (\alpha\alpha)$.
The bound-state energies for the two- and three-body systems are given by $E_{\alpha\alpha}\equiv -B_{\alpha\alpha}$, 
$E_{\alpha\beta}\equiv -B_{\alpha\beta}$ and $E_3=-B_3$, respectively; with the energy of the $s-$wave elastic colliding 
particle given by $E_k$. 
In the following, we first recover the bound-state three-body formalism, restricted to the $s-$wave case when all the sub-systems 
being bound. Next, by introducing the appropriate boundary conditions we  extend the formalism to atom-dimer collision.

\subsection{Three-body $\alpha\alpha\beta$ bound-state}
The bound-state coupled equation for separable potentials is usually written in terms of the spectator functions 
for the particles $\alpha$ and $\beta$, given by  $\chi_{\alpha}({\bf q};E_3)$ $\chi_{\alpha}({\bf q};E_3)$.
 For $s-$wave, this coupled equation is given by
\begin{eqnarray}\label{eq:1}
 \chi_{\alpha}(q)&=&\tau_{\alpha}(q;E_3)\int_0^{\infty} dk k^2 \left[K_2(q,k;E_3)
 \chi_{\alpha}(k)
 \right.\nonumber\\ &+&\left. 
 K_1(q,k;E_3)\chi_{\beta}(k)\right]\\
 \chi_{\beta}(q)&=&2\tau_{\beta}(q;E_3)\int_0^{\infty}dk k^2 K_1(k,q;E_3)
 \chi_{\alpha}(k),\nonumber
\end{eqnarray}
where $\chi_{\alpha}(q)\equiv \chi_{\alpha}(q;E_3)$ and $\chi_{\beta}(q)\equiv\chi_{\beta}(q;E_3)$.
 $\tau_{\alpha}$ and $\tau_{\beta}$ are the respective two-body t-matrix for the $\alpha\beta$ and
$\alpha\alpha$ subsystems, with $K_1$ and $K_2$ being the appropriate kernels, which will be 
explicitly given in the following according to the kind of form-factors one considers for the two-body interactions.

By considering the definitions
\begin{eqnarray}\label{kas}
\frac{k_\alpha^2}{2\mu_{\alpha(\alpha\beta)}}&\equiv&E_3-E_{\alpha\beta},\;\; 
\frac{k_\beta^2}{2\mu_{\beta(\alpha\alpha)}}\equiv E_3-E_{\alpha\alpha},
\end{eqnarray}
 with $j=\alpha,\beta$,  $\tau_{j}$, $\chi_{j}$ and the coupled Eq.~(\ref{eq:1}) can be conveniently redefined.  
As both subsystems are bound, we have 
\begin{equation}\label{eq:7}
\tau_{j}(q;E_3)\equiv \frac{\bar{\tau}_{j}(q;E_3)}{q^2+|k_j^2|},\;\; 
\chi_{j}(q)\equiv \frac{h_{j}(q;E_3)}{q^2+|k_j^2|},
\end{equation}
with
\begin{eqnarray}\label{eq:9}
 h_{\alpha}(q;E_3)&=&\bar{\tau}_{\alpha}(q;E_3)\int_0^{\infty} dk k^2 \left[K_2(q,k;E_3){h_{\alpha}(k;E_3)
 \over(k^2+|k_\alpha^2|)}\right. \nonumber \\ 
 &+&\left. K_1(q,k;E_3){h_{\beta}(k;E_3)\over (k^2+|k_\beta^2|)}\right], \\
 h_{\beta}(q;E_3)&=&\bar{\tau}_{\beta}(q;E_3)\int_0^{\infty}dk k^2 K_1(k,q;E_3){h_{\alpha}(k;E_3)\over (k^2+|k_\alpha^2|)}
\nonumber .\end{eqnarray}
The expressions for $\bar{\tau}_j$ and kernels $K_{1,2}$ are given in the following subsection C, by considering the specific 
potential models that have we are using.

\subsection{Atom-dimer collision}
For the scattering of a particle $\alpha$  by the $\alpha\beta$ bound subsystem, we should first redefine the expression
for ${\tau}_\alpha$ given in Eq.~(\ref{eq:7})  (considering that $k_\alpha^2>0$), such that
${\tau}_\alpha(q;E_3)\equiv \bar{\tau}_\alpha(q;E_3)/(q^2-k_\alpha^2-{\rm i}\epsilon)$.  
Next, the formalism is extended to  obtain the scattering amplitude by introducing the required boundary condition.
For the $s-$wave, this condition is given by   
\begin{eqnarray}\label{eq:11}
 \chi_{\alpha}(q)\equiv2\pi^2\frac{\delta(q-k_i)}{q^2}+4\pi\frac{h_{\alpha}(q;E_3)}{q^2-k_\alpha^2-{\rm i}\epsilon}
,\end{eqnarray}
where $k_\alpha$ is given by Eq.~(\ref{kas}), with $E_3>0$ in this case.
So, the coupled equations (\ref{eq:9}) are replaced by
{\small
\begin{eqnarray}\label{eq:13}
&&h_{\alpha}(q;E_3)=\bar{\tau}_{\alpha}(q;E_3)\Bigg\{\frac{\pi}{2}K_2(q,k_\alpha;E_3)
+\int_0^{\infty} dk k^2\times\\
&&\hspace{-.0cm}\times\Bigg[K_2(q,k;E_3) {h_{\alpha}(k;E_3)\over (k^2-k_\alpha^2-{\rm i}\epsilon)}
 +K_1(q,k;E_3)\frac{h_{\beta}(k;E_3)}{q^2-k_\beta^2}\Bigg]\Bigg\},\nonumber\\
&& h_{\beta}(q;E_3)=\bar{\tau}_{\beta}(q;E_3)\Bigg\{\frac{\pi}{2}K_1(k_\alpha,q;E_3)\nonumber\\
&&+\int_0^{\infty}dk k^2 K_1(k,q;E_3) {h_{\alpha}(k;E_3)\over(k^2-k_\alpha^2-{\rm i}\epsilon)}\Bigg\}
.\end{eqnarray}
}
\subsection{Zero-range and finite-range interactions with corresponding kernels}
When using zero-range interactions, a momentum cut-off is required to regularize the formalism, within a 
renormalization procedure. For that, in the kernels a subtraction procedure is used with a regularizing
momentum parameter $\mu$, such that the kernels $K_{1,2}$ and $\bar{\tau}_{j}$ used in the formalism
are given by 
{\small \begin{eqnarray}\label{eq:2}
 K_{i=1,2}(q,k;E_3)&\equiv& G_i(q,k;E_3)-G_i(q,k,-\mu^2), \nonumber\\
  G_1(q,k;E_3)&=&\int_{-1}^{1}dx\left[E_3+{\rm i}\epsilon-\frac{q^2}{m}
- \frac{k^2}{2\mu_{\alpha\beta}}-\frac{kqx}{m}\right]^{-1}\nonumber\\
  G_2(q,k;E_3)&=&\int_{-1}^{1}dx\left[E_3+{\rm i}\epsilon-\frac{q^2+k^2}{2\mu_{\alpha\beta}}
 -\frac{kqx}{Am}\right]^{-1},
 \end{eqnarray}
 \begin{eqnarray}
 \bar{\tau}_{\alpha}(q;E_3)&\equiv& 
  \frac{\mu_{\alpha(\alpha\beta)}}{2\pi\mu_{\alpha\beta}^2}
 \left[\kappa_{\alpha\beta}+\kappa_{3,\alpha\beta}(E_3)
 \right],\\
 \bar{\tau}_{\beta}(q;E_3)&\equiv& 
 \frac{\mu_{\beta(\alpha\alpha)}}{2\pi\mu_{\alpha\alpha}^2}
 \left[\kappa_{\alpha\alpha}+\kappa_{3,\alpha\alpha}(E_3)
 \right] 
 ,\end{eqnarray}}
where
\begin{eqnarray}
   \kappa_{\alpha\alpha}&\equiv&\sqrt{-2\mu_{\alpha\alpha} E_{\alpha\alpha}},
   \;\;\; \kappa_{\alpha\beta}\equiv\sqrt{-2\mu_{\alpha\beta}
   E_{\alpha\beta}}\nonumber \\
   \kappa_{3\alpha\alpha}(E_3)&\equiv&\sqrt{-2\mu_{\alpha\alpha}
   \left[E_3-   \frac{q^2}{2\mu_{\beta(\alpha\alpha)}}
      \right]},\nonumber \\
   \kappa_{3\alpha\beta}(E_3)&\equiv&\sqrt{-2\mu_{\alpha\beta}
   \left[E_3-\frac{q^2}{2\mu_{\alpha(\alpha\beta)}
   }\right]}.
\end{eqnarray}
For finite-range interaction, we  assume a rank-one separable Yamaguchi potential, given by
\begin{eqnarray}
 V_{ij}(p,p')=\lambda_{ij}\left(\frac{1}{p^2+\gamma_{ij}^2}\right)\left(\frac{1}{p'^2+\gamma_{ij}^2}\right),
\end{eqnarray}
where $ij=\alpha\alpha$ or $\alpha\beta$, respectively, for the $\alpha\alpha$ or $\alpha\beta$ two-body subsystems.
$\lambda_{ij}$ and $\gamma_{ij}$ refer to the strength and range $r_{ij}$ of the respective two-body interactions.  
As in the present approach we consider only bound (negative) two-body subsystems, $E_{ij}=-B_{ij}$, the corresponding 
relations for the strengths and ranges are given by
\begin{eqnarray}
\lambda^{-1}_{ij}=\frac{-2\pi\mu_{ij}}{\gamma_{ij}(\gamma_{ij}+\kappa_{ij})^2},\;\;\;
 r_{ij}=\frac{1}{\gamma_{ij}}+\frac{2\gamma_{ij}}{(\gamma_{ij}+\kappa_{ij})^2}.
\end{eqnarray}

In this case, $K_{1,2}$ and $\bar{\tau}_j$ are given by the following:
{\small \begin{eqnarray}
 K_1(q,k;E_3)&=&\int_{-1}^1 dx \left[q^2+\frac{k^2}{4}+qkx+\gamma_{\alpha\alpha}^2\right]^{-1}\\
 &\times&\left[k^2+\frac{q^2A^2}{(A+1)^2}+\frac{2qkAx}{(A+1)}+\gamma_{\alpha\beta}^2\right]^{-1}\nonumber \\
 &\times&\left[E_3+{\rm i}\epsilon-\frac{q^2}{m}-\frac{k^2}{2\mu_{\alpha\beta}}-\frac{qkx}{m}\right]^{-1},\nonumber
 \\
 K_2(q,k;E_3)&=&\int_{-1}^1\hspace{-0.2cm}dx\left[k^2+\frac{q^2}{(A+1)^2}+\frac{2qkx}{(A+1)}+\gamma_{\alpha\beta}^2\right]^{-1}\nonumber \\
 &\times& \left[q^2+\frac{k^2}{(A+1)^2}+\frac{2qkx}{(A+1)}+\gamma_{\alpha\beta}^2\right]^{-1}\\
 &\times&\left[E_3+{\rm i}\epsilon-\frac{(q^2+k^2)}{2\mu_{\alpha\beta}}-\frac{qkx}{Am}\right]^{-1},\nonumber
\end{eqnarray}}
\begin{eqnarray}
\bar{\tau}_{\alpha}(q;E_3)&\equiv& 
  \frac{\mu_{\alpha(\alpha\beta)}}{\pi\mu_{\alpha\beta}^2}\Bigg[
  \frac{\gamma_{\alpha\beta}(\gamma_{\alpha\beta}+\kappa_{\alpha\beta})^2}
{2\gamma_{\alpha\beta}+\kappa_{3\alpha\beta}(E_3)+\kappa_{\alpha\beta}}     \\
&\times&[\gamma_{\alpha\beta}+\kappa_{3\alpha\beta}(E_3)]^2
[\kappa_{\alpha\beta}+\kappa_{3\alpha\beta}(E_3)]\Bigg],
\nonumber \\
 \bar{\tau}_{\beta}(q;E_3)&\equiv&
   \frac{\mu_{\beta(\alpha\alpha)}}{\pi\mu_{\alpha\alpha}^2}\Bigg[\frac{
\gamma_{\alpha\alpha}(\gamma_{\alpha\alpha}+\kappa_{\alpha\alpha})^2
}
{2\gamma_{\alpha\alpha}+\kappa_{3\alpha\alpha}(E_3)+\kappa_{\alpha\alpha}}
\\
 &\times&
[\gamma_{\alpha\alpha}+\kappa_{3\alpha\alpha}(E_3)]^2 [\kappa_{\alpha\alpha}+\kappa_{3\alpha\alpha}(E_3)]\Bigg]
\nonumber.
\end{eqnarray}
In our approach, the  parameters of the separable interactions are fixed by the corresponding bound-state 
energies, as well as by the effective ranges (when considering finite-range interactions).

Finally, the scattering observables, $s-$wave phase shift $\delta_0$, cross-section $\sigma$, and 
absorption parameter $\eta$
are obtained by using the on-shell scattering amplitude $h_{\alpha}(k;E_3)$, considering that
\begin{eqnarray}
&&h_{\alpha}(k;E_3) =\frac{S_\alpha-1}{2\,{\rm i}\,k}\,\,,\, S_\alpha=\eta \,e^{2{\rm i}\delta_0 },
\label{sm} \\
&&\frac{d\sigma}{d\Omega}=|h_{\alpha}(k;E_3)|^2 \, ,
\end{eqnarray}
where $S_\alpha$ is the scattering matrix for the elastic $s-$wave channel and
$\eta\leq 1$ is the absorption parameter. 

\section{results}

In this section we present our main results and analysis
for the scattering of an atom $^4$He colliding with a weakly bound diatomic molecule composed 
by $^4$He with $ ^{6,7}$Li  or $^{23}$Na.
In this regard, by considering that the two-body subsystems in this study are weakly-bound, the relevant low-energy observables that we focus on  are the $s-$wave cross-sections, which 
are directly related to the $s-$wave phase-shifts $\delta_0$, and the corresponding absorption parameter. 
For that,  we use different two-body interactions, namely the
renormalized zero-range model and a finite-range model given by one-term separable Yamaguchi potential. In both
cases, we assume as inputs the available binding energies from different realistic model calculations. 
In the case of the ZR model,  the inputs are introduced in the renormalization procedure; whereas, for the finite-range case, the inputs are used to adjust the parameters (range and strength) of the Yamaguchi potential.  

\subsection{$^4$He$_2-^{7,6}$Li  Efimov molecules}

\begin{figure}[thb]
\begin{center}
\includegraphics[width=8cm,height=5.5cm]{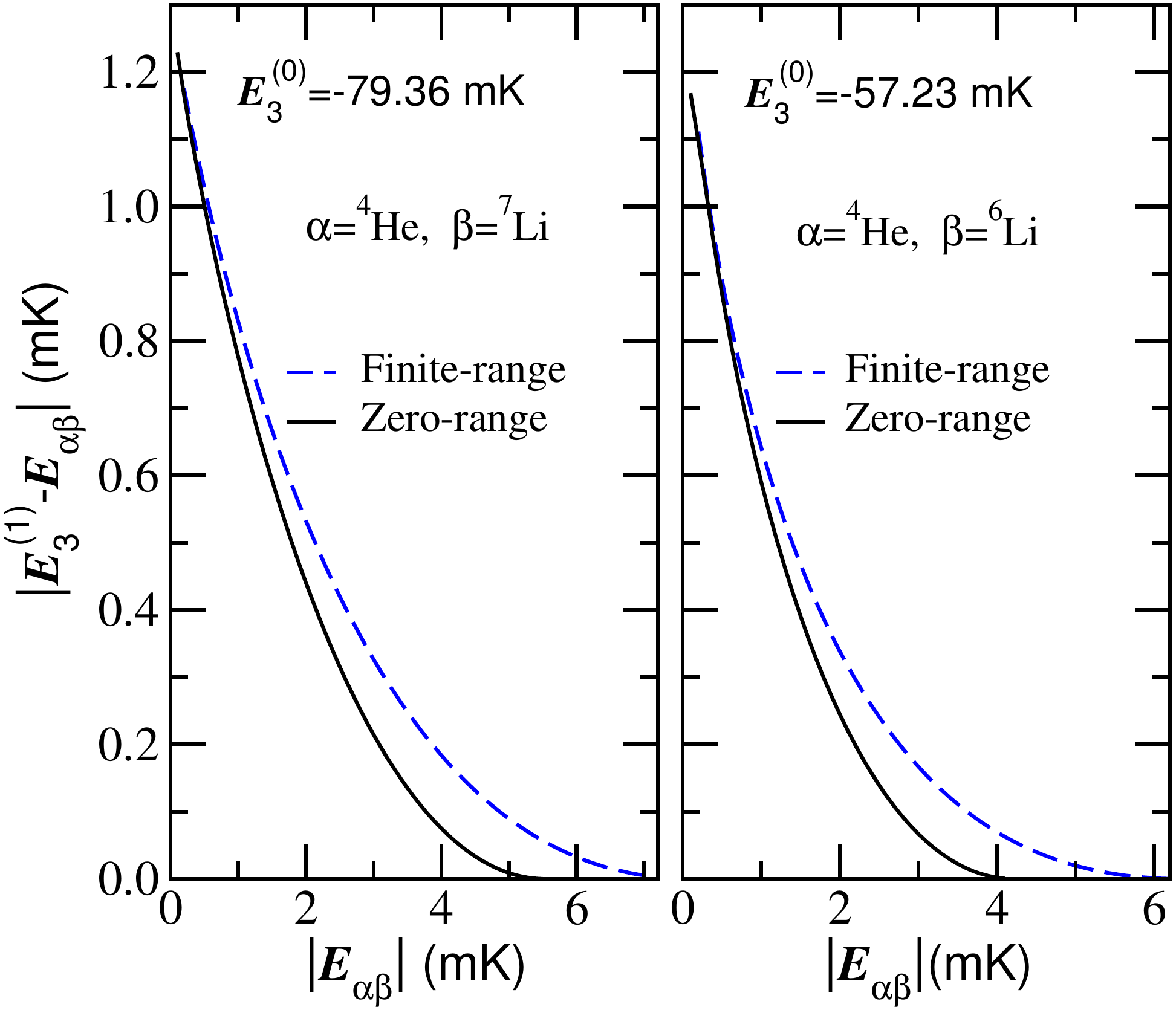}
\end{center}
\vspace{-.6cm}
\caption{For the $^4$He$_2-^7$Li (left frame) and  $^4$He$_2-^6$Li (right frame) three-body systems, with 
$\alpha\equiv^4$He and $\beta\equiv^7$Li, $^6$Li, respectively, we show the behavior of the corresponding 
three-body excited states $E_3^{(1)}$, which are represented by the absolute value of 
$E_3^{(1)}-E_{\alpha\beta}$, in terms of the dimer binding energies $E_{\alpha\beta}$.
As indicated inside the frames, the results are obtained by using zero-range and finite-range two-body interactions, 
with the given values for $E_3^{(0)}$.
The $^4$He$_2$ bound state in both the cases is fixed to $B_{\alpha\alpha}=|E_{\alpha\alpha}|=1.31$mK. }  
\label{fig1}
\end{figure}

Before moving to the main focus of this presentation on the atom-molecule scattering, we study the relevance 
of the range in the  formation of excited Efimov triatomic states by comparing results obtained with both potential models, 
in situations where such states are expected to exist. 
These are the cases of $^4$He$_2-^7$Li and $^4$He$_2-^6$Li, 
where we fix the well-known  $^4$He$_2$ dimer energy, $B_{\alpha\alpha}=$1.31 mK, together with the corresponding 
ground-state three-body energies 
given in Ref.~\cite{2017Suno}:  $E_3^{(0)}=-$79.36 mK for $^4$He$_2-^7$Li; and $-$57.23 mK for  $^4$He$_2-^6$Li.
Our results for  the excited three-body bound-state energies (reduced by the corresponding 
two-body bound-state energies), obtained by the ZR and Yamaguchi models, are shown in the two frames of Fig.~\ref{fig1} 
as a function of the two-body binding energies. 

As shown in the left frame of Fig.~\ref{fig1}, the finite-range
 Yamaguchi potential, which reproduces the given $^4$He$_2$ dimer and 
$^4$He$_2-^7$Li ground-state binding energies, will allow an excited Efimov state if we have $^4$He$-^7$Li dimer 
bound with binding energy less than $\sim$7 mK. Correspondingly, as shown in the right frame, the upper limit
of the  $^4$He$-^6$Li dimer energy to produce an excited three-body state is $\sim$6 mK, when using the FR 
Yamaguchi potential.
For the zero-range model, the upper limit for the binding energy of the
dimer to allow an excited state is $\sim$5.5 mK for $^4$He$_2-^7$Li; being $\sim$4.1 mK  for $^4$He$_2-^6$Li.
 
 In Fig. \ref{fig1} are shown results for particular examples, considering the given binding energies, of the universal 
 scaling behavior theoretically found for weakly-bound triatomic states when considering 
 two-species atomic systems close to the Efimov limit, where the sizes of the ground state trimer and dimers are 
 much larger than the interaction range. Such situation is associated with a large probability of occupation 
 of the classically forbidden region dominated by the dynamics of free Hamiltonian, scale invariant and model independent. 
 The correlation between the excited triatomic binding energy and the ground state comes from the
 breaking of the continuous scale invariance to a discrete one, which translates in a universal scaling function as 
 the limit cycle of the discrete Efimov scaling~\cite{1999-Amorim,1999-Bedaque}, when the range of the interaction is driven towards 
 zero (see, e.g., the reviews \cite{2012-review,2017-Hammer}).

The interaction range allows more room to the formation of the Efimov state, namely the 
critical value of the $^4$He-$^{6,7}$Li molecular binding can be somewhat larger, as one can
see  in Fig.  \ref{fig1} through the comparison between the ZR and Yamaguchi potential results. 
The effective range expansion
says that the scattering length for a given dimer binding energy increases with the effective range as
\begin{equation}\label{aab}
a_{\alpha\beta}\approx \left(\kappa_{\alpha\beta}-\frac12 r_{\alpha\beta}\kappa_{\alpha\beta}^2\right)^{-1}
\approx \kappa_{\alpha\beta}^{-1}+\frac12 r_{\alpha\beta}\kappa_{\alpha\beta}\, ,
\end{equation}
 which shows that the cut of the tail of the attractive Efimov long-range potential 
  should increase with the effective range. Therefore, the formation of the large triatomic excited state 
  is favoured when the range of the short interaction increases, in the situation where  ground state
   energy is kept fixed.

 The scaling plot shown in the figure was first derived and presented in Fig.2 of Ref.~\cite{1999-Amorim} for trimers
composed by identical bosonic atoms. A general study of the
universal three-particle behavior, with two kind of particles, was previously presented in Ref.~\cite{1997-Amorim}.
We complement the plots shown in Fig.~\ref{fig1} with Table~\ref{tab1}, where realistic values for the two-body 
subsystem (given in the 2nd and 3rd columns) and for the three-body ground-state energy (4th column) are shown
from Refs.~\cite{1998Yuan} (also considered in Ref.~\cite{2000Delfino}) and \cite{2017Suno}. In the second block of the 
table we have the corresponding predicted  three-body excited states, with the values obtained in Ref.~\cite{2017Suno} 
given in the 5th column. 
Our corresponding results, when using the two- and three-body binding energies given in the 2nd and 4th columns, are 
presented in the 6th and 7th columns, by using the zero-range and finite-range approaches. In all these
cases, the binding energy of the $^4$He is $B_{\alpha\alpha}=1.31$mK, with the corresponding scattering length
being $a_{\alpha\alpha}=100$\AA.

\begin{table}[thb!]
\caption{
For the three-body molecular systems identified in the first column, given the two-body energies and scattering
lengths in the 2nd and 3rd columns and the three-body ground-state energies in the 4th column, as given
in Ref.~\cite{2017Suno}, we have the first excited bound-state energies in the 5th to 7th columns. In the  5th column 
the results are from Ref.~\cite{2017Suno}. Our results for the excited states, using zero-range (ZR) 
and finite-range one-term Yamaguchi (FR) interactions, are shown in the 6th and 7th columns. In our notation, 
$\alpha$ and $\beta$ are identifying, respectively, the $^4$He and the other atomic species ($^{6,7}$Li, $^{23}$Na).
In all the cases, for the $^4$He dimer, we have the well-known value $B_{\alpha\alpha}=1.31$, with the scattering
length being $a_{\alpha\alpha}=$100\AA. 
}
{\footnotesize \begin{tabular}{c|ccc|ccc}
 \hline\hline
$\alpha\alpha\beta$&
$B_{\alpha\beta}$&
$a_{\alpha\beta}$&
$B_{3}^{(0)}$&
$B^{(1)}_{\alpha\alpha\beta}$&
$B^{(1)}_{3(ZR)}$ &
$B^{(1)}_{3(FR)}$ 
\\
$\alpha=^4$He&(mK)&(\AA)&(mK)&(mK)&(mK)&(mK)\\
\hline\hline
$\beta=^7$Li&5.622&48.84&79.36     &5.642&-&5.672\\
$\beta=^6$Li &1.515&100&57.23        &1.937&1.901&1.977\\
 \hline\hline
  \end{tabular}
  }
  \label{tab1}
 \end{table}

We have to add that the effect of the range in the case of  the molecule $^4$He$_2-\,^7$Li with the
parameters from ~\cite{2017Suno} and given in Table \ref{tab1} allows one Efimov excited state with binding
energy of  5.7 mK. In this case as the binding energy of $^4$He$-\,^7$Li molecule is comparatively large with
respect to the ground state energy, the range gives the crucial contribution to increase the scattering length, and the
cut in the long-range effective Efimov potential, such that the excited state is barely bound. This state heals 
over quite incredible large distances, namely of about 800-900\AA. If that comes true, the the binding energy
of these excited state  will be a sensitive indirect measure of the interaction range. 
The other lithium isotope, $^6$Li,  forms a weakly bound molecule with 
 $^4$He and there is little effect of the interaction range in the $^4$He$_2-\,^6$Li Efimov excited state.

\subsection{ Elastic scattering of $^4$He on $^4$He-($^{6,7}$Li,$^{23}$Na) }

In order to pursue our aim in studying the atom-dimer systems with
 $\alpha\equiv ^4$He as the projectile and dimers
$\alpha\beta$, where $\beta\equiv ^7$Li, $^6$Li, and $^{23}$Na, in the next we provide the Tables~\ref{tab2} and 
\ref{tab3}, which we have considered to calculate the corresponding elastic atom-dimer $s-$wave cross-sections.

In Table~\ref{tab2}, we present available two- and three-body ground-state binding energies 
(absolute values, given in mK), obtained from different realistic potential models, (a1) to (a8), for the atomic system we are 
studying with $\alpha=^4$He, $\beta=^7$Li,$^6$Li and $^{23}$N.  
Specifically, (a1) is from \cite{2017Suno}; (a2) from \cite{1998Yuan}, with interactions from \cite{1996KTTY};
(a3) from \cite{2002Gianturco}, with potentials from \cite{1997Kleinekathofer,1999Kleinekathofer};
(a4) from \cite{2002Gianturco}, with potentials from \cite{1995Tang}, for $\alpha\alpha$ and \cite{1999Kleinekathofer}, 
for $\alpha\beta$;
(a5) from \cite{2000Gianturco}, with potentials from \cite{1991Aziz,1994Cveko};
(a6) from \cite{2002Gianturco}, with potentials from \cite{1994Cveko};
(a7) from \cite{2014Wu}, with potentials from \cite{1991Aziz}, for $\alpha\alpha$ and \cite{1999Kleinekathofer},
for $\alpha\beta$;
(a8) from \cite{2017Kolganova}.
These energies are used to adjust the
parameters of our zero-range and finite-range (FR) separable interactions. 

\begin{table}[tbh!]
\caption{Available two- and three-body ground-state binding energies (absolute values, given in mK), for the three 
atomic systems given by $\alpha=^4$He, $\beta=^7$Li, $^6$Li and $^{23}$Na, from different model potentials. 
(a1) from \cite{2017Suno};
(a2) from \cite{1998Yuan};
(a3), (a4) and (a6) from \cite{2002Gianturco};
(a5) from \cite{2000Gianturco};
(a7) from \cite{2014Wu};
(a8) from \cite{2017Kolganova}.
These data are being considered as inputs in our numerical approach on the atom-dimer collision. 
}
{\footnotesize \begin{tabular}{l||llllllll}
 \hline\hline
           &(a1) 	&(a2) 	&(a3)	&(a4)	&(a5)     	&(a6)  &(a7)	&(a8)\\
 \hline\hline
 $^4$He-$^7$Li 	&5.622	&2.16	&5.621	&5.621	&2.81	&2.81	  &5.355	&5.621\\
  $^4$He-$^{6}$Li	&1.515	&0.12	&-	&-	&0.33		&-	  &-		&1.515\\
 $^4$He-$^{23}$Na	&28.98	&28.98	&28.98	&28.98	&-		&-	  &-		&-\\
\hline
 $^4$He$_2-^{7}$Li	&79.36	&45.7	&65.6	&80.0	&73.4	&57.1 &78.73	&50.89\\
 $^4$He$_2-^{6}$Li	&57.23	&31.4	&-		&-		&51.94	&-	  &-		&35.45\\
 $^4$He$_2-^{23}$Na&150.9	&103.1	&148.5	&119.3	&-		&-	  &-		&-\\
 \hline\hline
  \end{tabular}
  }
  \label{tab2}
 \end{table}
 
\begin{table}[tbh!]
\caption{Parameters used in the separable interactions, with the corresponding ranges and scattering lengths, 
considered for the $^4$He-$^7$Li (upper part), $^4$He-$^6$Li (middle part) and $^4$He-$^{23}$Na (lower part). 
The references (first columns) are identified in the caption of Table ~\ref{tab2}.
For the $^4$He dimer, to fit the binding energy 1.31 mK and corresponding scattering length $a_{\alpha\alpha}$=100 \AA,
we have $\gamma_{\alpha\alpha}$=0.39 \AA$^{-1}$ and $r_{\alpha\alpha}$=7.34 \AA.
}
\begin{tabular}{c|ccc} 
\hline\hline
\multicolumn{4}{c} {\bf$^4$He-$^7$Li}\\
\hline\hline
references&
$\gamma_{\alpha\beta}$(\AA$^{-1}$)&
$r_{\alpha\beta}$(\AA)&
$a_{\alpha\beta}$(\AA)\\
\hline\hline
(a1)& 0.17& 14.77& 50.08\\
(a2)& 0.14 &       19.02 &    77.43\\
(a3)& 0.144 &       17.19 &      51.89\\
(a4)& 0.17 &       14.68 &      50.01\\
(a5)& 0.19   &     13.95 &    66.10\\ 
(a6)& 0.16 &       16.82 &    67.98\\ 
(a7)& 0.17 &      14.72&     51.01\\
(a8)& 0.11 &     21.04&    55.02 \\ 
 \hline\hline
   \multicolumn{4}{c} {\bf$^4$He-$^{6}$Li}\\
 \hline\hline
 (a1) & 0.17&   15.85&    90.38\\
(a2) & 0.14&    20.04&       300.37\\
(a5)&  0.19&  15.11&    182.77\\
(a8)&  0.12&      22.18&      94.40 \\
 \hline\hline
   \multicolumn{4}{c} {\bf$^4$He-$^{23}$Na}\\
 \hline\hline
(a1)& 0.16  &      12.44   &     25.34\\   
(a2)& 0.09 &  19.0  &  34.24\\
(a3)& 0.16 & 12.65 & 25.58  \\ 
(a4)& 0.11 & 15.99  & 29.80 \\
 \hline\hline
  \end{tabular}
  \label{tab3}
 \end{table}

For the case of FR, the 
parameters with corresponding ranges and scattering lengths, are shown in Table~\ref{tab3}, given in three blocks 
for the cases with $^4$He-$^7$Li, $^4$He-$^6$Li and $^4$He-$^{23}$Na. We observe that, in all the cases, for the
dimer $^4$He$_2$ binding energy, the accepted value $B_{\alpha\alpha}=$1.31 mK is being considered, with the
corresponding parameters given in the caption of this table.

\subsubsection{Exploring  parameter dependence}
 
\begin{figure*}[htb]
\begin{center}
\includegraphics[scale=0.4]{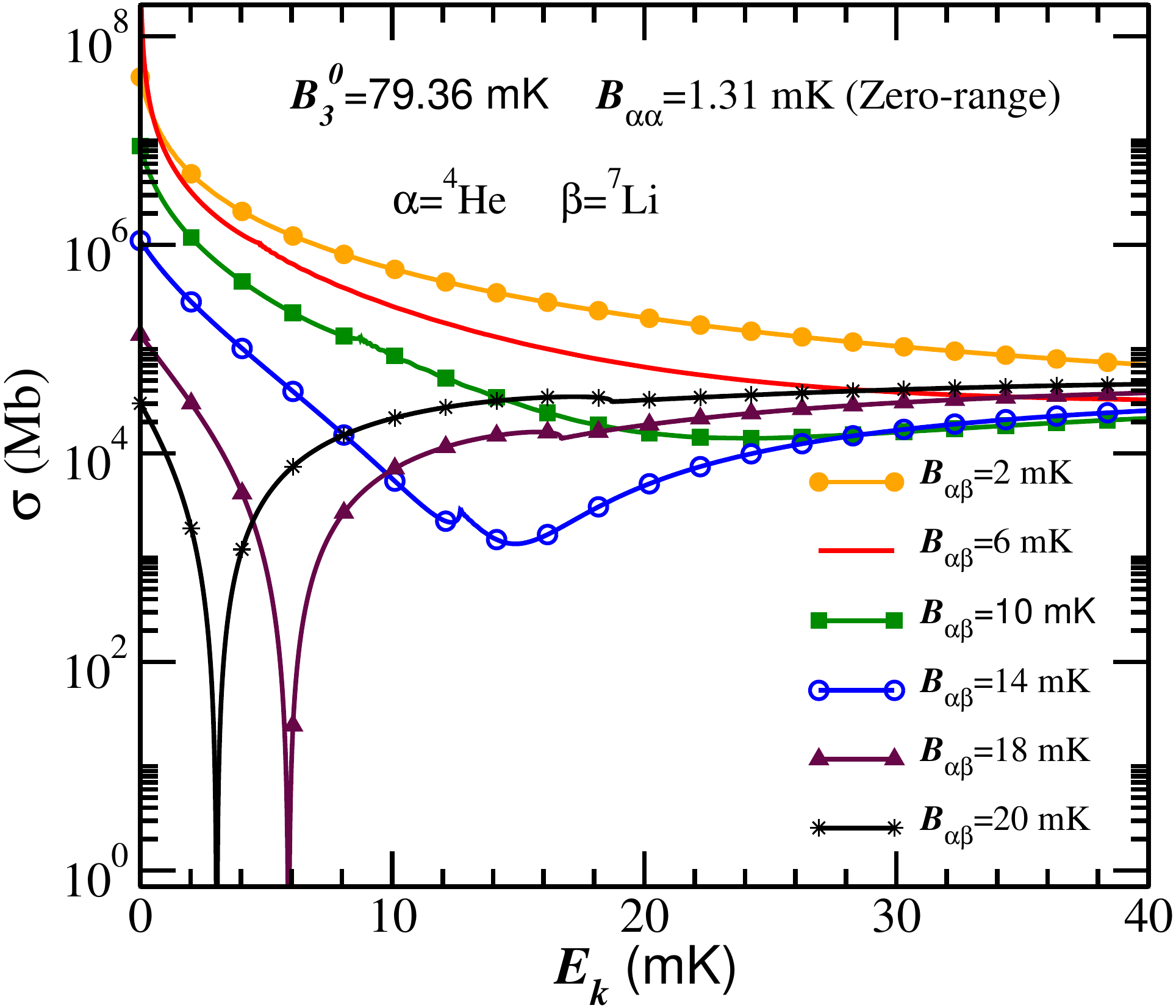}
\includegraphics[scale=0.4]{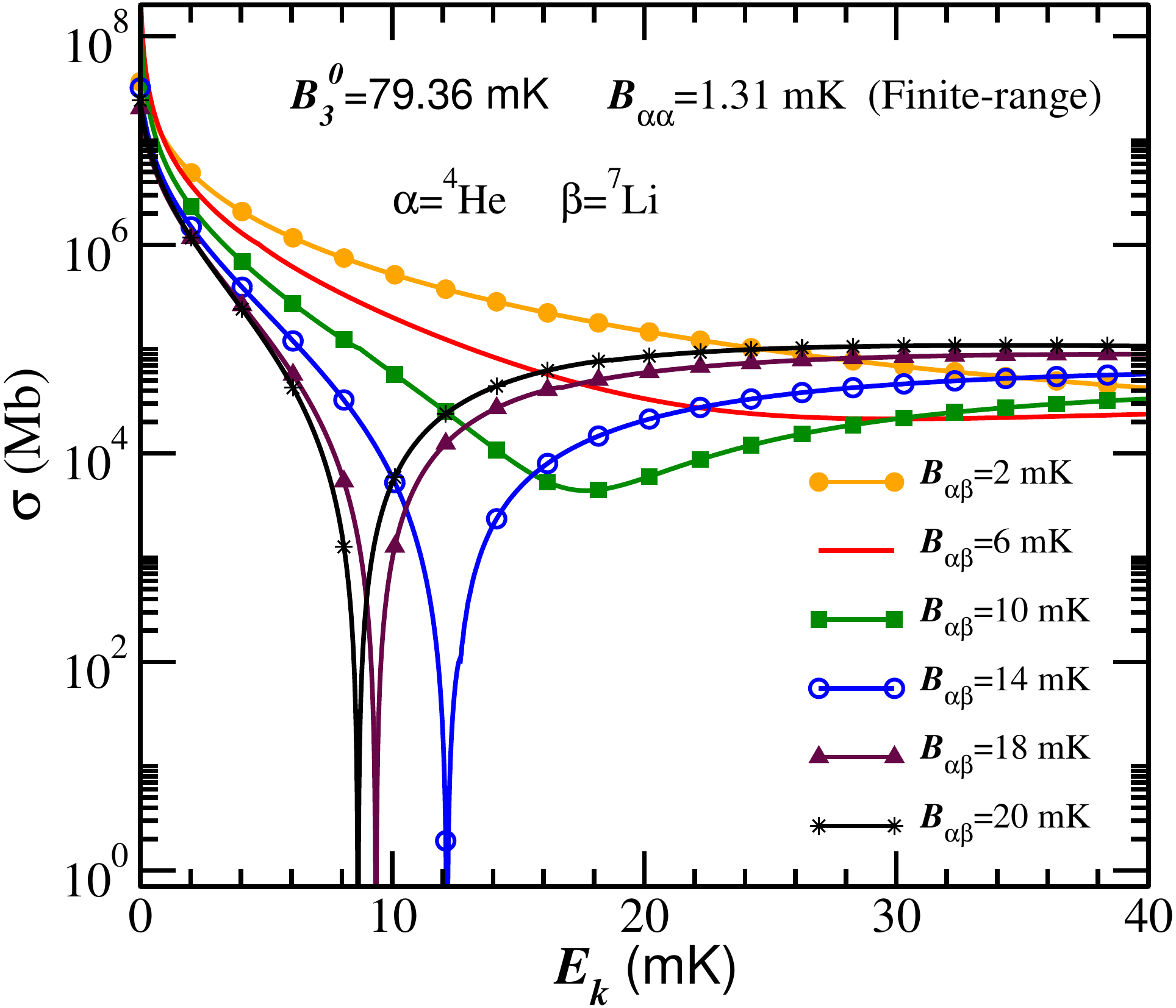}
\includegraphics[scale=0.4]{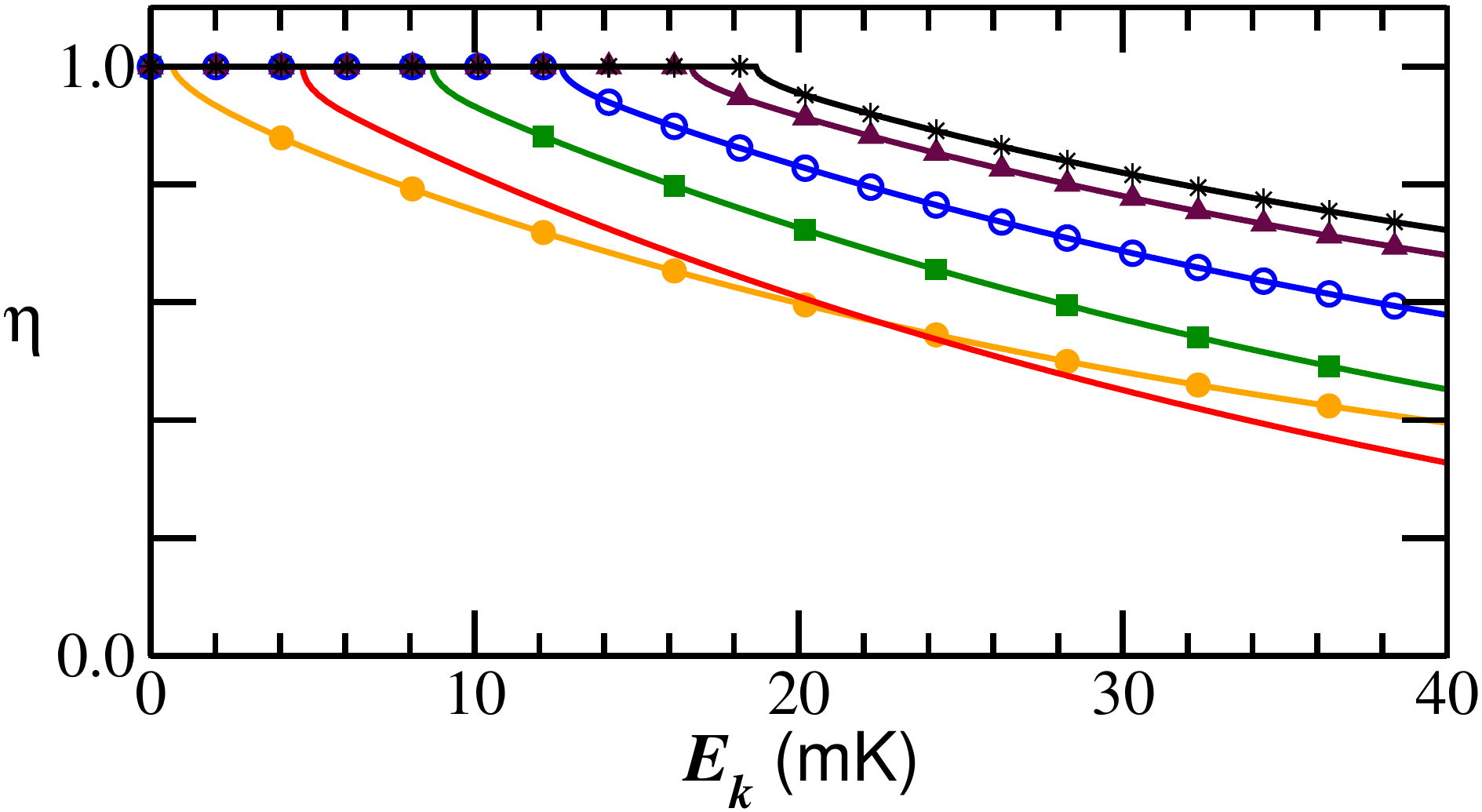}
\includegraphics[scale=0.4]{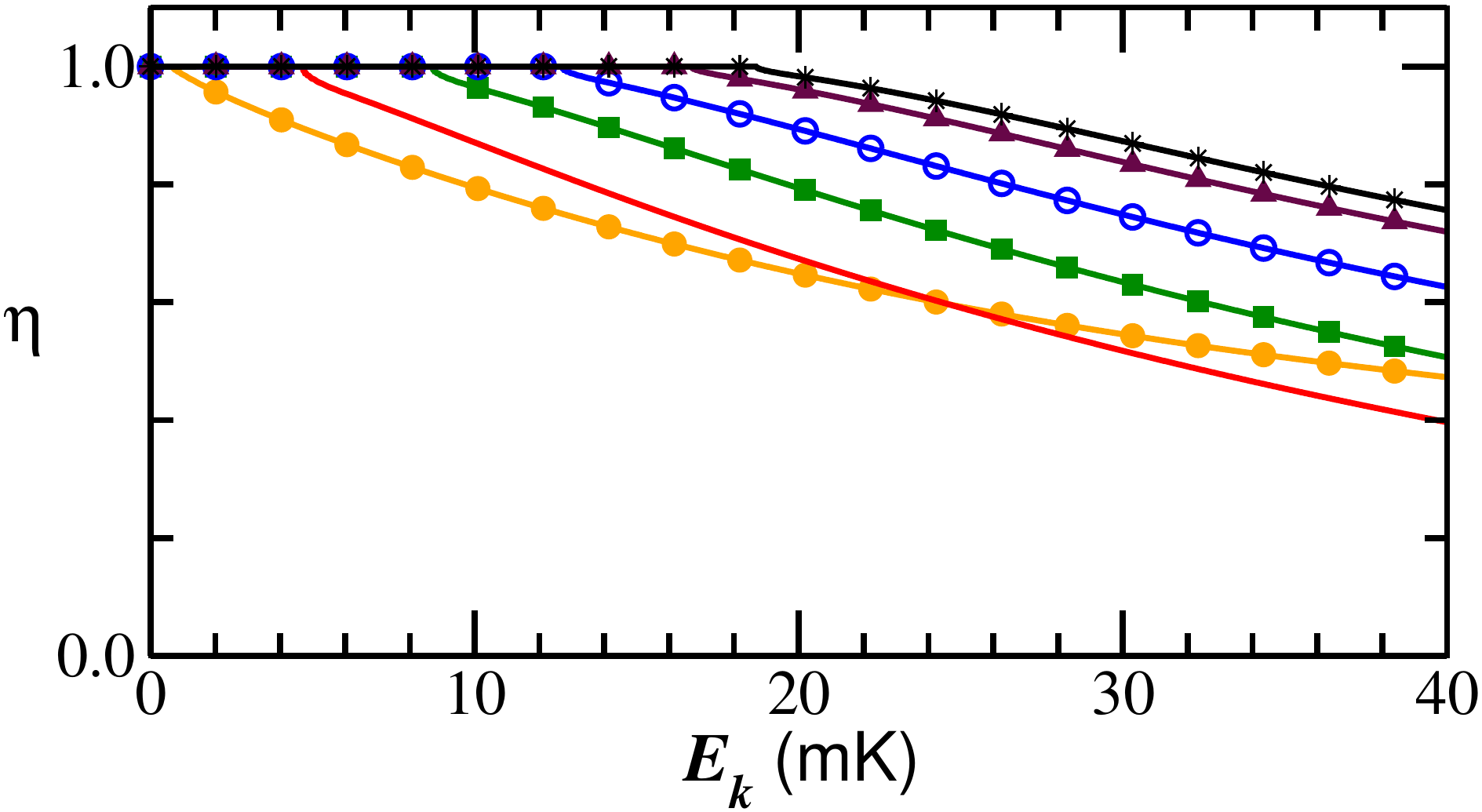}
\end{center}
\vspace{-.6cm}
\caption{The $s-$wave cross section $\sigma$ (upper frames), with the corresponding absorption parameters 
$\eta$ (lower frames), for the scattering of $\alpha\equiv^4$He by the $\alpha\beta$ ($^4$He$-^7$Li)
 system as a function of the kinetic energy $E_k$ of the projectile in the center-of-mass system. The results are 
 given by using zero-range potential in the left frames; and by using Yamaguchi separable potentials in the right frames.
 In our parametrization, the binding energies of the subsystem $^4$He$_2$ and three-body ground-state are, 
 respectively, fixed to $B_{\alpha\alpha}=$1.31 mK and $B_{3}=$79.36 mK, considering several binding energies 
for the subsystem $\alpha\beta$, as given inside the frames.}  
\label{fig2}
\end{figure*}

 In the present study on scattering observables for the elastic channel of an atom and diatomic molecule collision,
 we start by presenting some general results 
 when considering that the binding energy for the $\alpha\beta$ subsystem can be arbitrarily varied, keeping fixed the 
 other two- and three-body binding energies. To explore the general features of this parameter dependence, 
 both in the case of the ZR and Yamaguchi models,
 we  use the example of the atom $\alpha\equiv ^4$He  colliding 
 elastically with  the dimer $(\alpha\beta)\equiv$ ($^4$He$-^7$Li ). 
 The results,  obtained by using zero-range and finite-range one-term separable 
 interactions, are shown for the $s-$wave cross-sections and corresponding absorption parameters, respectively in 
the upper and lower panels of Fig.~\ref{fig2}, as functions of the collision energy $E_k$ in the rest frame.

The comparison between the ZR and FR results  in Fig.~\ref{fig2} shows quite similar results when 
the two dimer binding energies are comparable, such that  $B_{\alpha\beta}\lesssim 5 B_{\alpha\alpha}$. 
However, as expected the interaction range starts to be more relevant for larger values of $B_{\alpha\beta}$.
The present results are evidencing that, as we increase $B_{\alpha\beta}$  for a fixed ground state 
triatomic molecular binding energy, a minimum starts to emerge in $\sigma$, 
which have the tendency to move towards  some value of $E_k$ as  $B_{\alpha\beta}$ increases.
This behavior is quite clear when using finite-range interactions, as the range parameters 
 are more relevant to obtain correctly the scattering observables.  Possibly such curious property is due to the
less efficient role of the decreasing $a_{\alpha\beta}$ in cutting the long range potential as compared to 
the larger $a_{\alpha\alpha}$.

We should also noticed a cusp  in the plots at energies $E_k=B_{\alpha\beta}-B_{\alpha\alpha}$, 
corresponding to the position where the new channel is open. For  $E_k>B_{\alpha\beta}-B_{\alpha\alpha}$,
we are verifying the effect of the absorption, as shown in the lower panels, where  
we notice that $\eta$ tends to saturate with the energy. 
This is clearly shown in the case that $B_{\alpha\beta}=$2 mK, implying that for $E_k\gg B_{\alpha\beta}$ 
there is no more possibility to increase the absorption.

The comparison between the results of ZR and Yamaguchi models in Fig. \ref{fig2} for the 
$^4$He-($^4$He$-^7$Li ) $s-$wave cross-section show more less cases of 
minima for the ZR calculations. This curious effect can already be thought as being reasonable, because  
when the effective range is considered  $a_{\alpha\beta}$ increases for a given $B_{\alpha\beta}$
(c.f. Eq. \ref{aab}), and therefore there is more room for the log-periodic behavior of the wave function 
establish a zero in $\delta_0$ for the Yamaguchi potential  when compared  to the ZR model. 
Notice that, the zero turns to a minimum if above the threshold to open the rearangement 
channel, which we notice by the cusp for energies below the minimum.

The appearance of zeros in the elastic $s-$wave cross sections is
traced back to the dominance of the log-periodic behavior of the scattering 
wave function inside the long-range Efimov potential, extensively discussed in ~\cite{2018Shalchi}. 
Of course as the scattering lengths moves to 
larger values more cycles of the wave function appears in the Efimov potential is possible, 
allowing the presence of zeros in the cross-sections and the maxima. However, on the other side this phenomena
concentrates on small values of the kinetic energies, as the opening of a scattering channel tends to washout these 
minima, as the probability flux is driven to new open channels.

\subsubsection{Realistic $^4$He-$^7$Li and  $^4$He-$^6$Li parameters }

\begin{figure}[thb]
\begin{center}
\includegraphics[width=8cm,height=6cm]{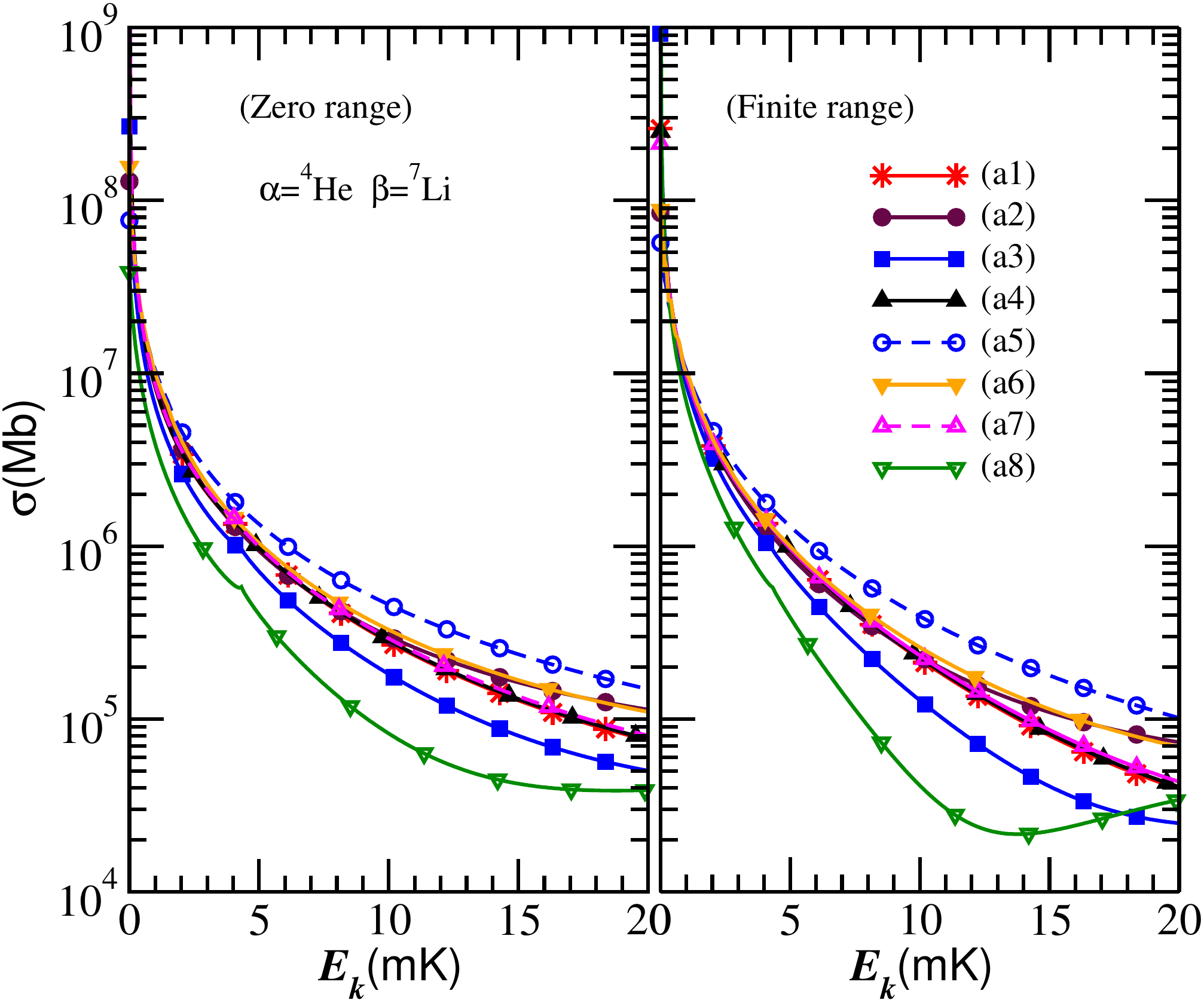}
\includegraphics[width=8cm,height=6cm]{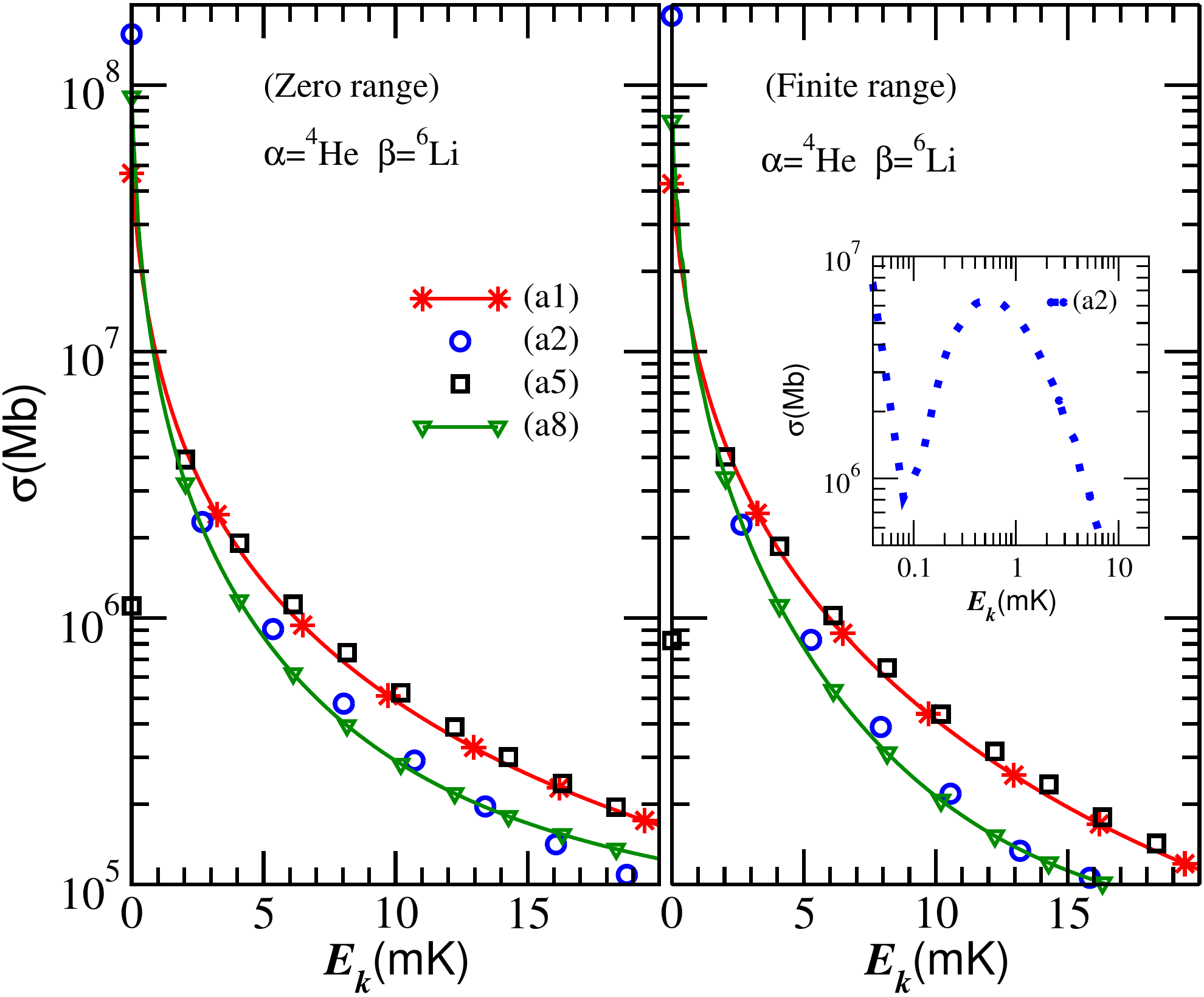}
\end{center}
\vspace{-.6cm}
\caption{ Results obtained for the $s-$wave cross-section, for the scattering of $^4$He from the
dimers $^4$He-$^7$Li (upper panels) and $^4$He-$^6$Li (lower panels). In the left panels we have
the results by using zero-range interactions; with finite-range results being presented in the right
panels. In both the cases,  
we use binding energies obtained from different realistic model calculations, as indicated (inside 
the right upper panel for $\beta=^7$Li and inside the left panel for $\beta=^6$Li)  
with the corresponding references given in the caption of Fig.~\ref{fig1}.
For the model (a2), we also show the results by an inset in the lower-right panel.
The finite-range-interaction parameters are given in Table~\ref{tab3}.
}  \label{fig3}
\end{figure}

By using different realistic model inputs for the $^4$He-$^7$Li and  $^4$He-$^6$Li dimer binding energies, as given 
in Table~\ref{tab2},  the results for the cross-sections are shown in Fig.~\ref{fig3}, respectively, in the upper and lower panels. 
In both the cases, we consider zero-range (left panels) and finite-range (right panels) interactions which are fitting the 
respective binding energies presented in Table~\ref{tab2}.  

We should observe the characteristic behavior of the plots in Fig.~\ref{fig3}
when the collision energy $E_k$ is very small approaching to zero. For the case that the two-body 
binding energies for for $^4$He-$^6$Li are very small, as the ones provided by the models from Refs.~\cite{1998Yuan} and \cite{2000Gianturco} 
[identified by (a2) and (a5), respectively] , we notice that each curve of the cross-sections are presenting 
a maximum for $E_k<1$~mK. 
Such behavior can better be understood by scaling all the energies ($E_k$ and the two-body binding energies), 
using the corresponding three-body ground-state energies, as it was done in Ref.~\cite{2018Shalchi}. As 
learned from the studies for atom-molecule collision at very low-energies performed 
in Ref.~\cite{2018Shalchi}, when considering small
enough values for the sub-system binding, as $E_k$ is decreased, one should observe
maxima and minima in the corresponding $s-$wave cross-section.

This behavior can be seen in the two cases that we use as inputs dimer 
energies $B_{\alpha\beta}$ very small in comparison with the three-body ground-state energies.
Indeed, when the two-body energy are close to the unitary limit, the cross-section should present a 
series of maxima and minima, for enough small values of $E_k$,  in the limit that the mass ratio $m_\alpha/m_\beta$ 
becomes very small, with similar behavior as the Efimov excited states (see~\cite{2018Shalchi}).
However, in our present case, no more than one maximum is observed in each curve, because the mass-ratios are 
not enough small as the ones considered in Ref.~\cite{2018Shalchi}. Therefore, the curious behavior observed in 
two plots shown in the lower panels of Fig.~\ref{fig3} (when using $^6$Li) is a manifestation of the same singular 
behavior for the scattering function $k\cot\delta_0$, long-time known from neutron-deuteron studies~\cite{Fuda} 
and recently studied in Ref.~\cite{2018Shalchi}. 

We call the reader attention to the minimum in the cross section produced by the input parameters of the model 
(a2) with the Yamaguchi potential in the case of the elastic collision of $^4$He on $^4$He-$^6$Li molecule as 
shown by the inset in the lower panel of Fig. \ref{fig3}. As verified in Ref.~\cite{2018Shalchi}, when going to a
limit with very small two-body binding, at some specific energies the scattering observable $k\cot\delta_0$ 
turns out to be singular, leading to zeros in the corresponding cross-section. The zero will happen if there is 
no absorption, which is the case for $E_k<B_{\alpha\beta}-B_{\alpha\alpha}$.  However, 
in the present case of the model (a2), $B_{\alpha\beta}-B_{\alpha\alpha}<0$, such that $E_k$ can never be 
less than zero. 
As absorption is always possible, instead of a zero we observe a minimum in the $s-$wave cross-section, which
follows from Eq.~(\ref{sm}),
\begin{equation}
\sigma= \pi \frac{|\eta e^{2{\rm i}\delta_0}-1|^2}{k^2},
\end{equation}
such that $\sigma= \pi |\eta -1|^2/k^2$ for $\delta_0=0$, characterising a minimum instead of a 
zero when $\eta\ne 1$.
A similar behavior can be seen with the results for $\sigma$ given by the model (a5). 
However, as in this case
$B_{\alpha\beta}=0.33$ mK is not so small as the value we have from model (a2), the minimum is
not clearly characterized in the results of Fig.~\ref{fig3}, but evidenced by a point very close to
$E_k=0$ (see in both lower panels of Fig.~\ref{fig3} the results represented with black squares).  
About the possible observation of a zero or minimum in the $s-$wave cross-section for the scattering of $^4$He in 
$^4$He-$^6$Li, it should be disregarded as not being expected from more recent realistic calculations identified 
by (a1) and (a8) (from Refs.~\cite{2017Suno} and \cite{2017Kolganova}, respectively).

For the case of  the $^4$He-$^7$Li molecule, as shown in the upper panels of Fig.~\ref{fig3} 
 with different models, most of the results for the cross section have similar behaviors, 
 considering that the energies $B_{\alpha\beta}$ are not so small in comparison with the three-body ones.  
 The model (a8), given by Ref.\cite{2017Kolganova}, which is showing  
a minimum in the cross-section for $E_k$ near 13 mK, is indicating the minimum of the cross-sections 
for larger collision energies, as the ratio $B_{\alpha\beta}/B_{3}^{(0)}$ is increased as already discussed when exploring
the cross-section for different inputs in Fig.~\ref{fig2}.
Among the models considered with $^7$Li, (a8) is the one which provides the larger value for the ratio 
$B_{\alpha\beta}/B_{3}^{(0)}$. As a general remark, we notice that the curves for the cross-sections,
for $E_k\lesssim$ 15 mK  follows the energy ratios $B_{\alpha\beta}/B_{3}^{(0)}$, with the curves 
in the upper part being the ones with smaller values for this ratio.

\begin{figure}[thb]
\begin{center}
\includegraphics[width=8.5cm]{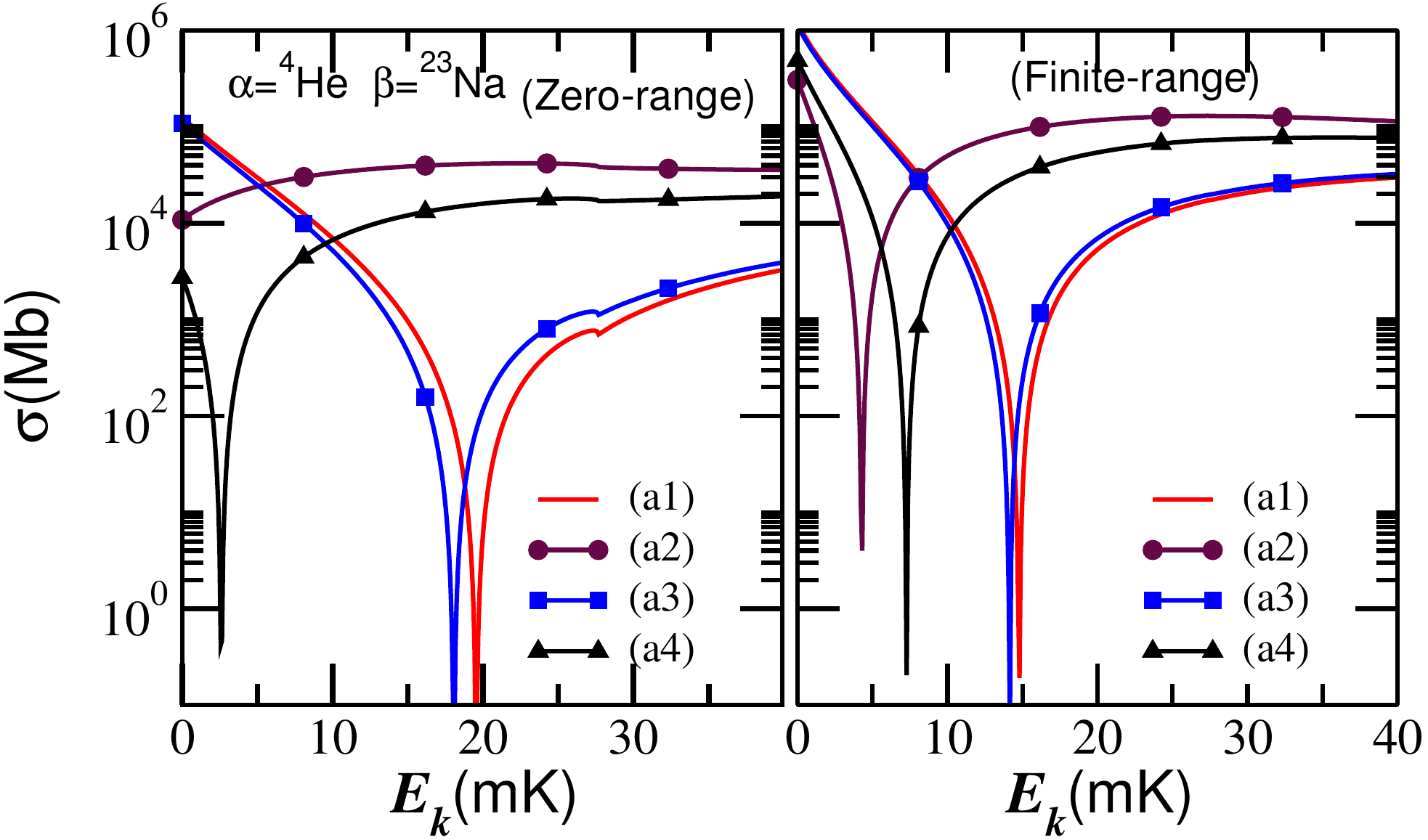}
\includegraphics[width=8.5cm]{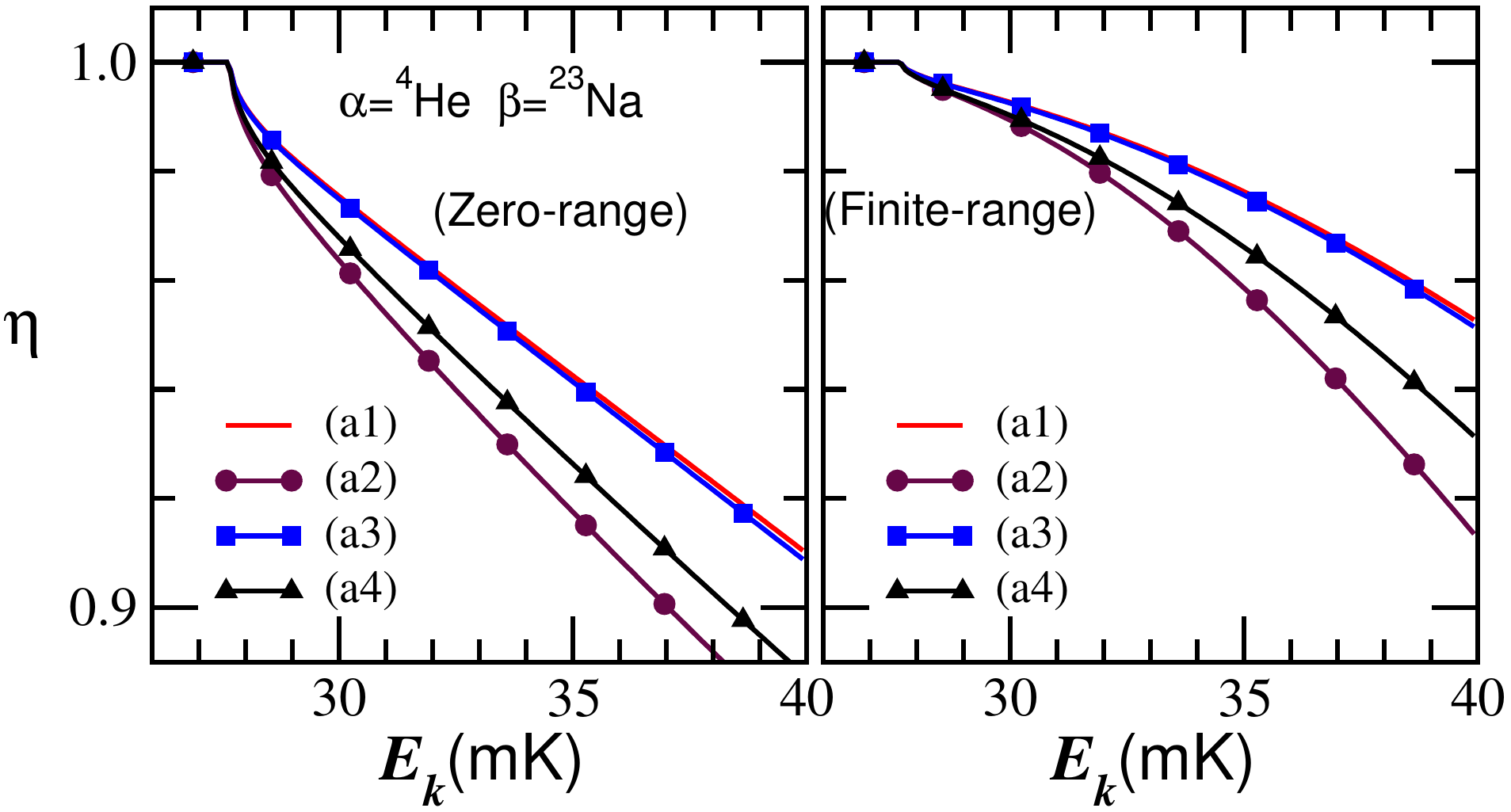}
\end{center}
\vspace{-.6cm}
\caption{ Cross-section $\sigma$ (upper panels) and absorption parameter $\eta$ (lower panels) $s-$wave results, 
for the collision of $^4$He in the $^4$He-$^{23}$Na dimer. The two and three-body energies used in the calculations
are indicated inside the panels, being  given by models quoted in the caption of Table \ref{tab2}.
The zero-range results are in the left frames; with the finite-range separable Yamaguchi results shown in the 
right frames.
}  \label{fig4}
\end{figure}

\subsection{Zero of $^4$He-($^4$He-$^{23}$Na) $s-$wave cross-section}

In case of the scattering of  $^4$He by  the  $^4$He-$^{23}$Na molecule,
 with four different realistic model calculations
available, our results are shown in Fig.~\ref{fig4} for the cross-sections (upper frames) and absorption parameters
(lower frames). 

We observe the same general features of the elastic $s-$wave cross-section already pointed out by the 
results shown in Fig.~\ref{fig2}, namely, with the increase of  the binding energy of the subsystem 
$\alpha\beta$ a minimum emerges, at some value of $E_k$. 
As $B_{\alpha\beta}$ increases, the tendency of this minimum seems to converge to some value of
$E_k$. This can be seen by matching the binding energies given for  $^4$He-$^{23}$Na in the last line
of Table~\ref{tab2} with the minima appearing in Fig.~\ref{fig4} when considering finite-range results.
Also, as noticed in the case of Fig.~\ref{fig2}, the convergence of the minima to some value $E_k$, when using 
ZR results is not so fast and clear as in case with FR results. 

More relevant to eventually future scattering experiments of  the $^4$He 
collision  with $^4$He-$^{23}$Na molecule is our conclusion of a minimum in the elastic $s-$wave
cross-section from finite-range interactions within our approach, and
using different realistic model calculations as inputs. Considering the more recent  realistic 
model calculations (a1) reported 
in Ref.~\cite{2017Suno}, a minimum should occur in the cross-section at a center-of-mass collision energy 
close to $E_k\sim$ 15 mK. This prediction seems robust as the the ZR model with the same 
input predict the position of the zero around  20 mK.

\section {Conclusion} 

We predict the $s-$wave scattering properties of cold  $^4$He  elastic collision with
 $^4$He$-^{6,7}$Li and $^4$He$-^{23}$Na molecules for center-of-mass kinetic energies up to 40 mK. 
Of particular experimental relevance, considering the actual investigations in cold atom laboratories,
we show the presence of a minimum in the $s-$wave elastic cross-section for the $^4$He$\to(^4$He-$^{23}$Na) 
scattering. This prediction was based on  calculations performed using finite-range separable interactions, 
where we used recent realistic model results for the molecular bound state energies as inputs to
get the model parameters.
By using the the binding energies reported in Ref.~\cite{2017Suno}, we predict that the elastic $s-$wave 
cross-section should have a minimum at a center-of-mass colliding energy close to $E_k\sim$ 15 mK.
In our approach, we have also obtained the corresponding $s-$wave absorption  parameter, which is 
relevant for defining the $s-$wave scattering amplitude. 

To access the importance of the range corrections to the phase-shifts and absorption parameters,
our calculations were performed with the zero-range model and also a finite-range one-term separable potential.
The results present some sensitivity to the potential range when the binding energies for the two-body 
subsystems are not small enough with respect to the three-body ground-state energy. 
It is well-known that close to the Efimov limit, namely zero dimer binding energies, 
the low energy three-body observables  are model independent and dominated by few low energy
scales, as in our case the diatomic and ground state triatomic binding energies. 

The model independence is exemplified in this work with universal scaling plots, considering the correlation
of the excited state energy of the $^4$He$_2$-$^{6,7}$Li with the $^4$He-$^{6,7}$Li molecule binding.
Such correlations are pronounced close to the unitary limit, however  
our examples of cold collisions are not at the unitarity taking into account the realistic
potential model results for the  binding energies of the di- and triatomic molecules and atom-atom 
scattering lengths. Despite of that we have shown that while elastic $s-$wave cross-section presents 
some sensitivity on the potential range  the basic universal predictions are not destroyed,  
 like the robust presence of the zero in the elastic $s$-wave cross section  of 
$^4$He on $^4$He$-^{23}$Na molecule, which we expect to motivate experimentalist to observe
such a property at the root of the universal Efimov physics. 

Finally, we should mention some perspective on further related investigations.
First, as stated in the introduction, a direct extension of this work is to explore contribution of higher partial 
waves in the total cross-section, which are expected to be non-negligible as we move to higher energies.
Also relevant, in our understanding, are the possible inelastic processes in the atom-dimer $\alpha\to(\alpha\beta)$
collision, such as three-body rearrangements going to $(\alpha\alpha)+\beta$ or total dissociation
($\alpha+\alpha+\beta$); processes expected to be significant at the collision energies we have used, 
deserving a separate detailed investigation.

{\bf Acknowledgments}
This work was partially supported by Funda\c {c}\~{a}o de Amparo \`{a} Pesquisa do Estado de S\~{a}o Paulo
(grant \#17/05660-0), Coordena\c c\~ao de Aperfei\c coamento de Pessoal de N\'\i vel Superior, Conselho 
Nacional de Desenvolvimento Cient\'{\i}fico e Tecnol\'{o}gico and Project INCT-FNA  (CNPq grant \# 464898/2014-5).

\end{document}